\documentclass[final,5p,times,twocolumn,sort&compress]{elsarticle}

\usepackage{amssymb}
\usepackage{amsmath}
\usepackage{subcaption}

\usepackage{placeins}

\journal{Physics of the Dark Universe}

\pdfoutput=1 

\usepackage[T1]{fontenc} 

\usepackage{graphicx,epsfig}
\usepackage{xspace}
\usepackage{verbatim}
\usepackage{textcomp}
\usepackage[utf8]{inputenc}

\newcommand{\mtlep}{\ensuremath{m_{\mathrm T}^{\ell}}}
\newcommand{\pt}{\ensuremath{p_{\mathrm T}}}
\def\etmiss{\ensuremath{E_{\mathrm{T}}^{\mathrm{miss}}}\xspace}

\def\ptmiss{\ensuremath{\vec p^{\mathrm{\ miss}}_\mathrm{T}\xspace}}
\def\ptl{\ensuremath{\vec p^{\mathrm{\ \ell}}_\mathrm{T}\xspace}}

\newcommand{\mttwo}{\ensuremath{m_{\mathrm{T2}}}\xspace}
\newcommand{\amttwo}{\ensuremath{am_{\mathrm{T2}}}\xspace}

\def\ttbar{\ensuremath{t\bar{t}}}

\newcommand{\bjet}{\ensuremath{b}-jet}
\newcommand{\btagged}{\ensuremath{b}-tagged}
\def\TeV{\ifmmode {\mathrm{\ Te\kern -0.1em V}}\else
                   \textrm{Te\kern -0.1em V}\fi}%
\def\GeV{\ifmmode {\mathrm{\ Ge\kern -0.1em V}}\else
                   \textrm{Ge\kern -0.1em V}\fi}%

\def \beq{\begin{equation}}
\def \eeq{\end{equation}}
\def \bea{\begin{eqnarray}}
\def \eea{\end{eqnarray}}

\usepackage{multicol}
\usepackage[colorlinks]{hyperref}

\begin{document}

\begin{frontmatter}



\title{Dark matter production in association with a single top-quark at the LHC
in a two-Higgs-doublet model with a pseudoscalar mediator}


\author[1]{Priscilla Pani}
\author[2]{and Giacomo Polesello}


\address[1]{CERN, Experimental Physics Department, \\ CH-1211 Geneva
23, Switzerland}
\address[2]{INFN, Sezione di Pavia\\ Via Bassi 6, 27100 Pavia, Italy} 


\begin{abstract}
The sensitivity of the LHC experiments to the associated 
production of dark matter with a single top is studied in the framework
of an extension of the standard model featuring two Higgs doublets and
an additional pseudoscalar mediator. 
It is found that the experimental sensitivity 
is dominated by the on-shell production of a charged Higgs boson, when this assumes a mass below $1\,\TeV$. Dedicated selections considering one and two lepton final states are developed to assess the coverage in parameter space for this signature
at a centre-of-mass energy of  $14\,\TeV$ assuming an integrated luminosity of 300~fb$^{-1}$.
For a pseudoscalar mediator $a$ with mass $150\,\GeV$ and maximally mixed with the 
pseudoscalar of the two Higgs doublets, values of  $\tan\beta$ up to 3 and down 
to 15 can be excluded at 95\%~CL, if the $H^{\pm}$ mass is in the range $300\,\GeV$-$1\,\TeV$.
This novel signature complements the parameter space coverage of the 
mono-Higgs, mono-$Z$ and $t{\bar t}$+$\etmiss$ signatures considered in previous publications 
for this model.
\end{abstract}

\end{frontmatter}

\section{Introduction}
\label{sec:introduction}
The nature of the dark matter (DM) is one of the key  open  questions of 
contemporary physics, and its experimental investigation is the subject of 
a worldwide effort based on several different and complementary 
experimental techniques.

The search for  particle DM produced at 
accelerators  is an essential part of this program, and it is
vigorously pursued at the CERN LHC, a proton-proton ($pp$) collider
currently operating at a center-of-mass energy of $13\,\TeV$.
Since the DM particles are weakly interacting 
they would escape the detector unseen when produced in $pp$ collisions
at the LHC.
The minimal experimental signature of DM production
at a hadron collider thus consists in events with a visible final-state
object $X$ recoiling against missing transverse energy
(\etmiss)  associated with the undetected DM. Based on the LHC data collected between
2009 and 2016, the ATLAS and CMS collaborations have
analysed a variety of such signatures involving jets of hadrons,
photons, electroweak (EW) gauge bosons, top and bottom quarks as well as
the Higgs boson in the final state~\cite{Aaboud:2016qgg,Sirunyan:2017hci,CMS-EXO-16-012,Aaboud:2017dor,Sirunyan:2017xgm,Sirunyan:2017ewk,Aaboud:2017uak,Aaboud:2017yqz,Aaboud:2017bja,Aaboud:2017rzf,Sirunyan:2017leh,Aaboud:2017phn,Aaboud:2017aeu}. Given the absence of 
a signal, upper limits on the production cross sections
have been obtained. The corresponding 
\etmiss searches have been  interpreted in the context of  three
different classes of theories: ultraviolet (UV) complete theories,
simplified models (see the reviews \cite{Abdallah:2014hon,Abdallah:2015ter, Abercrombie:2015wmb} for a complete list of references),
and effective field
theories~\cite{Beltran:2010ww,Goodman:2010yf,Bai:2010hh,Goodman:2010ku,Rajaraman:2011wf,Fox:2011pm} .
In particular, simplified models have become quite popular recently.
They allow the study of the different possible signatures
for DM production at the LHC focusing on the final state
kinematics, and thanks to their very limited set of parameters,
they provide a very  effective mapping of the phenomenological
space accessible to experimentation. 
While handy and in many cases useful, in general simplified models 
need to be employed with care. 
In some instances they might be too ``simplified'' to allow for an adequate
investigation of the experimental potential of DM searches,
as they sometimes neglect unique signatures  which may arise
from a more complete description of the interactions of DM
with the standard model (SM). 
In addition, it might happen that specific research channels become 
explicitly sensitive to the UV completion. Glaring examples are given 
by violation of unitarity and gauge invariance, which points
to the need for  more complex extensions of the SM
\cite{Chala:2015ama,Bell:2015sza,Kahlhoefer:2015bea,Bell:2015rdw,Haisch:2016usn,Englert:2016joy}.

Focusing on the cases where the interaction with DM is
mediated by a scalar or a pseudoscalar particle \cite{Cheung:2010zf, Buckley:2014fba, Haisch:2015ioa, Backovic:2015soa,  Arina:2016cqj,Banerjee:2017wxi}, a natural 
extension of the spin-0 simplified models is achieved by considering 
the mixing of the mediator with the Higgs boson. 
The experimental constraints on the Higgs boson couplings \cite{Khachatryan:2016vau}, 
however already severely constrain such a possibility.
One way to relax the constraints from Higgs physics is to add to the SM a second Higgs 
doublet (2HDM). \cite{Ipek:2014gua,No:2015xqa,Goncalves:2016iyg,Bauer:2017ota,Arcadi:2017wqi}. 
In this case the mediator that couples 
to DM can obtain its couplings to SM fermions from mixing with the 
second Higgs doublet. 

In the case of a 2HDM and a pseudoscalar mediator that couples to Dirac 
DM (2HDM+$a$), a detailed phenomenological analysis of the resulting  
\etmiss signatures at the LHC has been performed in~\cite{Bauer:2017ota}.
The conclusion drawn in that article is that the mono-Higgs and mono-$Z$
signatures provide a very good and complementary coverage of the parameter 
space of the model, with a minor but relevant role for the 
associated production of DM and a top-anti-top pair (DM$t\bar t$).
However the DM$t\bar t$ signature, as discussed in \cite{Haisch:2015ioa,Backovic:2015soa,Cheung:2010zf,Lin:2013sca,Buckley:2015ctj,Haisch:2016gry, Arina:2016cqj}, 
gives through the study of the kinematics of the top-anti-top pair, access 
to CP properties of the mediator and is therefore of great 
phenomenological interest in case of the future 
observation of a non-SM \etmiss signal.

A complementary signature with heavy quarks in the final state 
is the associated production of a single top quark with DM (DM$t$).
This signature has typically lower cross-section than DM$t\bar t$,
and has received little attention in the literature.   
A recent study \cite{Pinna:2017tay} based on a simplified model
with a singlet scalar or pseudoscalar mediator shows that the consideration of 
this process increases the coverage of 
existing analyses targeting the DM$t\bar t$ process.  
Given the promising result, it is worthwhile to  extend the investigation 
of \cite{Pinna:2017tay} in two directions. 
On the one hand it is necessary to check whether the DM$t$ signature 
is still promising in a more complete model that is not plagued by unitarity 
issues, as discussed above. We choose the 2HDM+$a$ model of 
\cite{Bauer:2017ota} as a benchmark model for this purpose.
On the other hand, the possible interest of the signature for future 
searches at the LHC can only be properly assessed if  a dedicated 
experimental analysis is developed, fully exploiting the final
state topology of the signal in order to suppress the SM backgrounds. 

The aim of this article is therefore to develop an experimental 
search strategy at the LHC for the DM$t$ signature, and to 
explore the parameter space of the chosen model that can be covered 
with the full LHC Run 3 statistics of  300~fb$^{-1}$ taken at a centre-of-mass 
energy of $14 \,\TeV$.

\section{The 2HDM+$a$ model}
\label{sec:2HDM}

\begin{figure}
\begin{center}
\begin{subfigure}{.23\textwidth}\centering
\includegraphics[width=\textwidth]{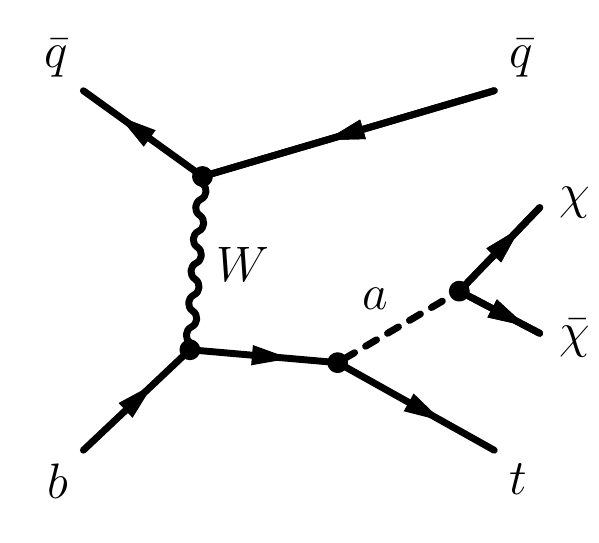}
\caption{}
\end{subfigure}
\begin{subfigure}{.23\textwidth}\centering
\includegraphics[width=\textwidth]{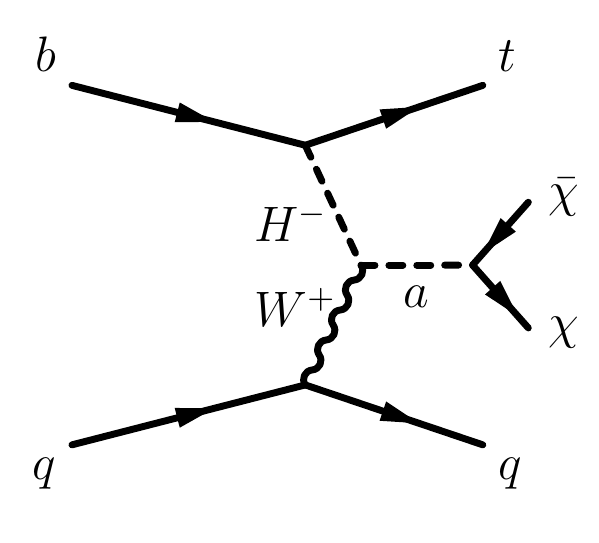}
\caption{}
\end{subfigure}
\caption{Representative diagrams for $t$-channel production of DM in association with a single top quark.}
\label{fig:feyn1}
\end{center}
\end{figure}

The extension to the SM  proposed in \cite{Bauer:2017ota} includes
a scalar sector with two Higgs doublets 
(see for example~\cite{Gunion:1989we,Branco:2011iw}), 
where the parameters relevant
for phenomenology are $\alpha$, the mixing angle of the two 
doublets and $\tan\beta$, the ratio of the vacuum expectation values (VEVs)
of the two doublets.
The angles $\alpha$ and $\beta$ are chosen according to the well-motivated alignment/decoupling limit of the~2HDM where $\alpha = \beta - \pi/2$. In this case $\sin \left ( \beta - \alpha \right ) = 1$ meaning that  the field~$h$ has SM-like EW~gauge boson couplings. It can therefore be identified with the boson of mass~$m(h) \simeq 125 \, \GeV$ discovered at the LHC \cite{Aad:2012tfa,Chatrchyan:2012xdj}. 

\begin{figure}
\begin{center}
\begin{subfigure}{.23\textwidth}\centering
\includegraphics[width=\textwidth]{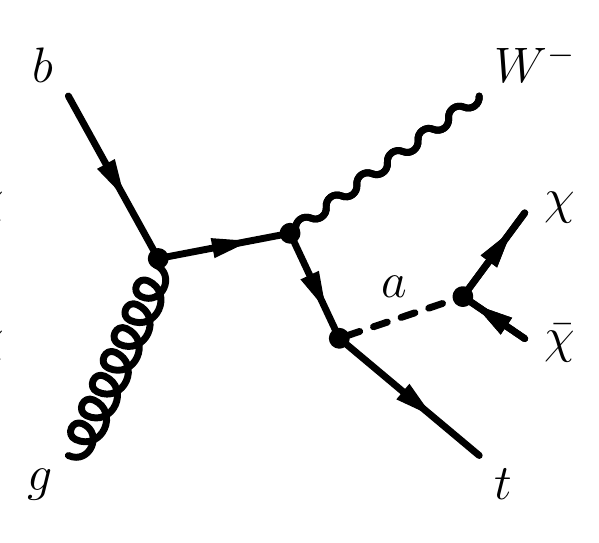}
\caption{}
\end{subfigure}
\begin{subfigure}{.23\textwidth}\centering
\includegraphics[width=\textwidth]{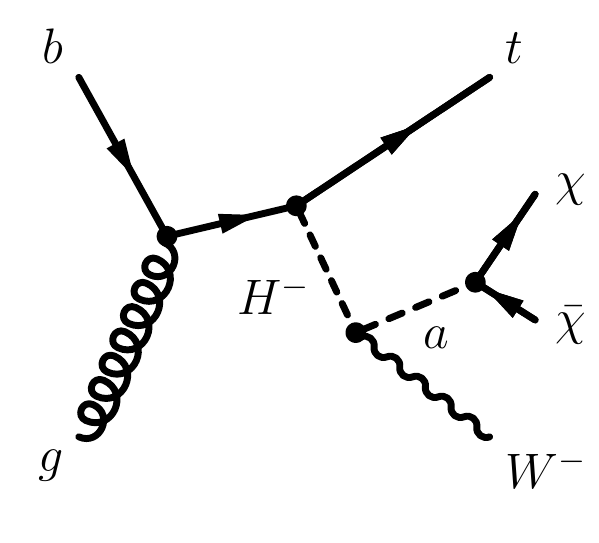}
\caption{}
\end{subfigure}
\caption{Representative diagrams for $tW$ production of DM
in association with a single top quark and a $W$ boson.}
\label{fig:feyn2}
\end{center}
\end{figure}

Dark matter is coupled to the SM 
by mixing a SU(2) singlet CP-odd mediator $P$ with the CP-odd Higgs that arises from the 2HDM potential.  The relevant interactions terms read
\beq \label{eq:VP} 
\begin{split}
V_{P}  & =  \frac{1}{2} \hspace{0.1mm} m_P^2  P^2 +  P \left ( i \hspace{0.05mm} b_P  \hspace{0.05mm}  H_1^\dagger H_2 + {\rm h.c.} \right ) \\[2mm]
& \phantom{xx} + P^2 \left (  \lambda_{P1}  \hspace{0.05mm}  H_1^\dagger H_1 +   \lambda_{P2}  \hspace{0.05mm}  H_2^\dagger H_2 \right )  \,,
\end{split}
\eeq
where $m_P$ and $b_P$ are parameters with dimensions of mass. The quartic 
portal interactions with couplings $\lambda_{P1}$ and $\lambda_{P2}$ do 
not affect the phenomenology studied in this paper, and 
$\lambda_{P1}$ and $\lambda_{P2}$  are thus set to zero hereafter.
The portal coupling $b_P$ appearing in~(\ref{eq:VP}) mixes the two neutral CP-odd weak 
eigenstates with $\theta$ representing the associated mixing angle
which emerges from the diagonalisation the mass-squared matrices 
of the scalar states. The resulting  CP-even mass eigenstates will be denoted by $h$ and~$H$, 
while in the~CP-odd sector the states will be called $A$ and $a$, where $a$  
denotes the mixing of the CP-odd  scalar from the 2HDM and of
the CP-odd mediator with weights $\sin\theta$ and $\cos\theta$, respectively. 
The scalar spectrum also contains two charged mass eigenstates $H^\pm$ of identical mass.

The Yukawa sector is built by respecting the so-called natural flavour 
conservation hypothesis, requiring that not more than one of the Higgs doublets 
couples to fermions of a given charge \cite{Glashow:1976nt,Paschos:1976ay}.
In the following we consider a 2HDM Yukawa assignment of type II 
yielding a coupling of the top quark (bottom quark and $\tau$ lepton) 
proportional to $-\cot\beta$ ($\tan\beta$) respectively.

The DM is taken to be a Dirac fermion  $\chi$ and is coupled to the 
pseudoscalar mediator $P$ through the interaction term
\beq \label{eq:Lx}
{\cal L}_\chi = - i \hspace{0.25mm} y_\chi P \hspace{0.25mm} \bar \chi \hspace{0.25mm} \gamma_5 \hspace{0.1mm} \chi \,.
\eeq
The DM coupling strength $y_\chi$ and the DM mass $m_\chi$ are further free parameters 
and are fixed as $y_\chi = 1$ and $m_\chi = 1 \, \GeV$ throughout our work.
The choice of the value of $m_\chi$ has no impact on the phenomenology addressed in this 
study as long as the decays $A,a\rightarrow\chi \bar \chi$ are kinematically open.

To avoid constraints from EW precision measurements, we furthermore 
assume that $m(H)=m(A)=m(H{^\pm})$. Together with
the restrictions specified above, this leaves a four-dimensional parameter space
including $\tan\beta$, $\sin\theta$, $m(H^{\pm})$ and $m(a)$ for the
phenomenological exploration in this paper.

\begin{figure*}
\begin{center}
\begin{subfigure}{.49\textwidth}\centering
\includegraphics[width=.9\textwidth]{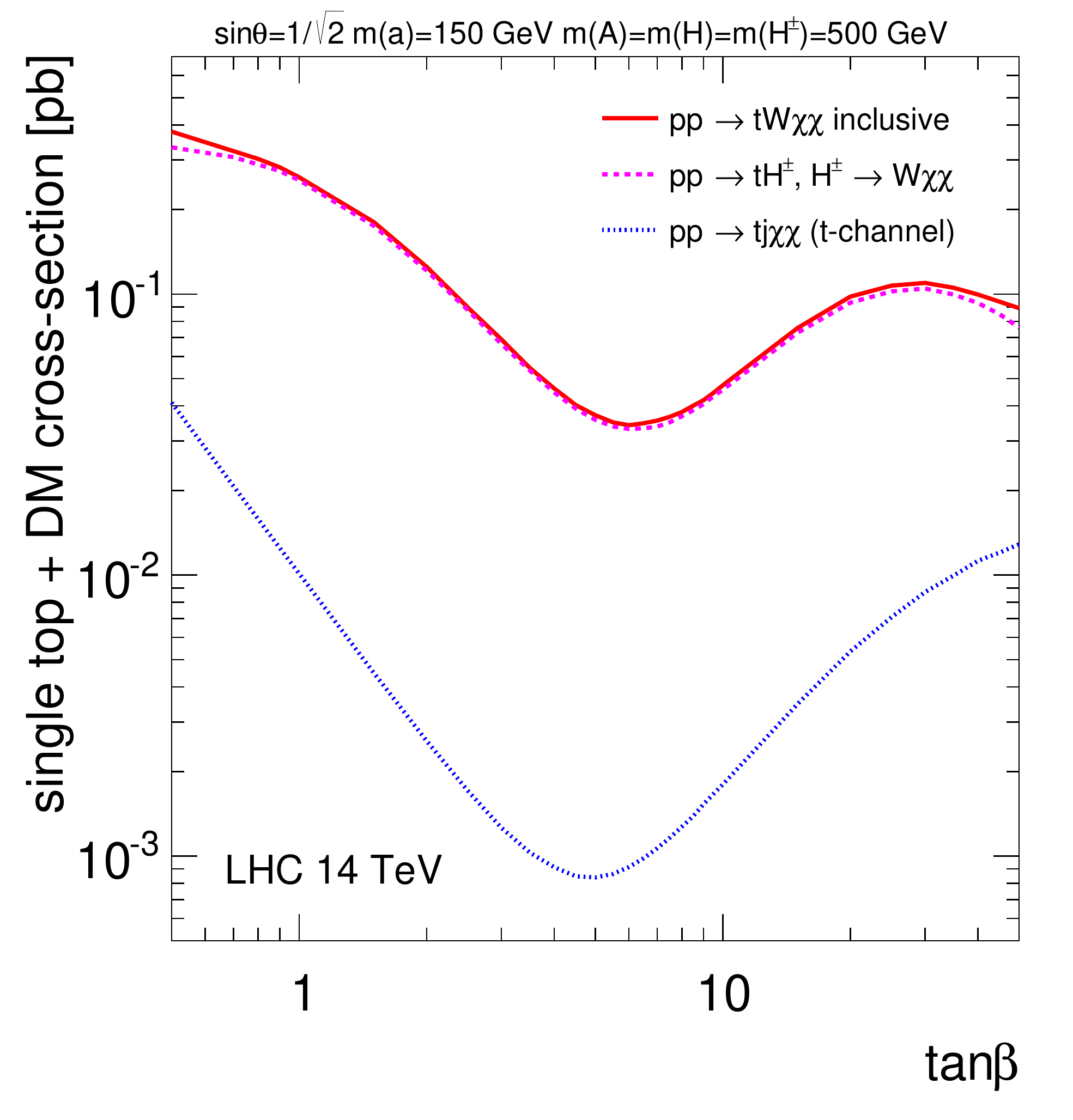}
\caption{}
\label{fig:xstba}
\end{subfigure}
\begin{subfigure}{.49\textwidth}\centering
\includegraphics[width=.9\textwidth]{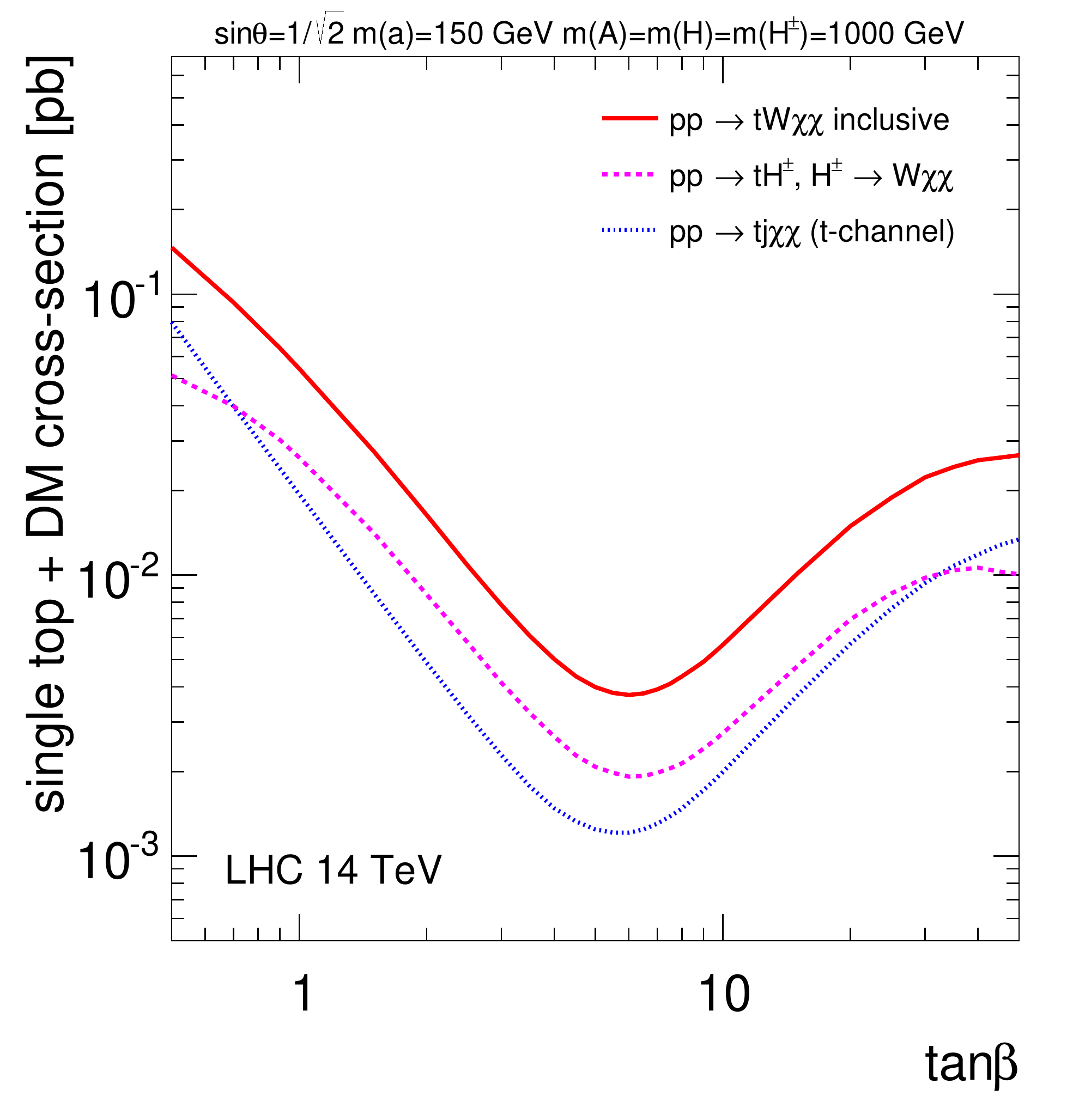}
\caption{}
\label{fig:xstbb}
\end{subfigure}
\caption{Cross-section for the associated production
of a top quark and DM for $pp$ collisions at $14 \; \TeV$
as a function of $\tan\beta$ for $m(a)=150\,\GeV$ and
$m(H^{\pm})=500\,\GeV$ (a) and $100\,\GeV$ (b). 
The full line 
corresponds to the $tW$ channel, while the
dotted line shows the result for $t$-channel production.
The dashed line indicates
the contribution to $tW$ production that arises from the on-shell production 
of a $H^{\pm}$ boson cascading into a $W^\pm$ and a DM pair.}
\label{fig:xstb}
\end{center}
\end{figure*}
\section{The DM$t$ signal}
Like single top production within the SM, the DM$t$ signature in the model~(\ref{eq:VP}) receives  three different types of contributions at leading order (LO) in QCD. These are $t$-channel production, $s$-channel production and
associated production together with a $W$ boson ($tW$).
The relative impact of the three production modes has been discussed in detail in~\cite{Pinna:2017tay} for the case of simplified spin-0 DM models.
DM$t$ production in the $s$-channel is, compared to the other channels, characterised by a very small cross-section, and we therefore neglect its contribution in our analysis.
The $t$-channel process $pp \rightarrow tj\chi \bar \chi$ receives the dominant contributions from the
two diagrams shown in Figure~\ref{fig:feyn1}. One has (a) the SM single top $t$-channel diagram
with radiation of the mediator from the top ($a$-strahlung), and (b)
the $t$-channel fusion
of a  charged Higgs and a $W$ into the mediator~$a$.
The two diagrams interfere destructively, and the amount of interference
decreases with increasing $H^{\pm}$ mass. As a result the $t$-channel production
cross-section in our model~(\ref{eq:VP}) is, for equivalent values of the mediator mass
and couplings, always smaller than the corresponding prediction
in the spin-0 DM simplified model. The observed destructive interference
ensures perturbative unitarity of the process $pp \rightarrow tj\chi \bar \chi$ in the 2HDM+$a$ model.

In the case of the $tW$ production channel it turns out that also two diagrams provide the dominant contributions to the DM$t$ cross-section. The
relevant graphs are shown in Figure~\ref{fig:feyn2}. The
$a$-strahlung diagram, also present in the simplified spin-0 DM model, is displayed on the left-hand side, while the right diagram
represents the associated production of a $H^{\pm}$ and a $t$ quark. Like in the case of  $t$-channel production the two diagrams interfere destructively
to ensure unitarity.
When the decay $H^{\pm}\rightarrow W^{\pm} a$ is possible, the $H^{\pm}$ is produced on-shell, 
and the cross-section of $pp \rightarrow tW\chi\chi$, 
assuming  $H^{\pm}$ masses of a few hundred \GeV, is around one order of magnitude larger 
than the one for the same process in the simplified model. Moreover the production 
and cascade decay of a resonance yields kinematic signatures
which can be exploited to separate the signal from the SM background. 
The dependence of the production cross-section on $\tan\beta$ 
for both the $t$-channel and $tW$ processes is shown in the two panels of 
Figure~\ref{fig:xstb}. Both panel employ 
$\sin\theta=1/\sqrt{2}$ and  $m(a)=150\,\GeV$, while
$m(H)=m(A)=m(H^{\pm})=500 \, \GeV$ and $1 \, \TeV$ is used in the left 
and right plot, respectively.
The cross-section for the contribution  of the on-shell production of 
$H^{\pm}$ to the $tW$ final state is also shown as a dashed line. The calculation is performed at LO in QCD in the 5-flavour scheme, 
and the Yukawa couplings of both $t$ and $b$ quarks are included in the calculation. 

From the shown results, one observes that the $tW$
contribution to the DM$t$ cross-section always dominates over the
$t$-channel, and that this dominance is more pronounced for lower
values of $m(H^{\pm})$. This feature is easy to understand by noting
that the $tW$ channel itself receives the dominant contribution from
 resonant $H^{\pm}$ production for charged Higgs masses of a few
 hundred \GeV, while for $m(H^{\pm})=1 \, \TeV$ resonant $H^{\pm}$
 production amounts to  only around 50\% of the $tW$ cross-section.

For all processes a  rapid decrease with increasing $\tan\beta$ is observed, 
with a minimum at $\tan\beta\simeq 5$, followed by a slower increase 
towards high $\tan\beta$ values.  The resonant $H^{\pm}$ production has a 
broad maximum for $\tan\beta$ in the range of $[20, 30]$.
This $\tan\beta$ dependence is the result of the interplay of
four factors: the production cross-section for $H^\pm$ production in $gb$ fusion  is
proportional to $m(t)^2\cot^2\beta + m(b)^2\tan^2\beta + {\rm const.}$;
the cross-section for diagrams where the $a$ is radiated off a top quark
is proportional to $\cot^2\beta$;
the branching ratio (BR) for $H^{+}\rightarrow W^{+}a$
acquires a $\tan\beta$  dependence from the
competition with  the decay $H^{+}\rightarrow t \bar b$;
finally, the BR for $a\rightarrow\chi \bar \chi$
decreases at high $\tan\beta$ since the  partial decay width  
$a\rightarrow b\bar{b}$ grows as $\tan^2 \beta$.

Since both the widths for $H^{\pm}\rightarrow W^{\pm}a$ and $a\rightarrow\chi\chi$ 
are proportional to $\sin^2\theta$, the cross-section for DM$t$ 
grows monotonically with $\sin\theta$.
For the following studies we fix the value of the mixing angle $\theta$ 
such that $\sin\theta=1/\sqrt{2}$, 
corresponding to maximal mixing in the pseudoscalar sector. 

\section{MC simulations}
\label{sec:montecarlo}
In this section we provide a brief description of the MC simulations used
to generate both the DM signal and the SM backgrounds and explain how muons, 
electrons, photons, jets and missing transverse energy, \etmiss, are built  in our detector simulation.
Throughout our analysis we will consider~$pp$ collisions at $\sqrt{s} =
14 \, \TeV$. 

\subsection{Signal generation}
\label{sec:signalgeneration}

The signal samples used in this paper are generated at LO using  the 2HDM+$a$
UFO model~\cite{Degrande:2011ua} implementation provided in~\cite{Bauer:2017ota}.
The DM$t$ events are generated with {\tt MadGraph5\_aMC@NLO}~\cite{Alwall:2014hca},
employing {\tt NNPDF3.0} parton distribution functions (PDFs)~\cite{Ball:2014uwa}.
The final-state top quarks
and $W$ bosons are decayed  with {\tt MadSpin}~\cite{Artoisenet:2012st}
and the events are showered with {\tt PYTHIA~8.2}
\cite{Sjostrand:2014zea} and a 5-flavour scheme is assumed.
We consider a grid in the ($m(H^{\pm})$, $\tan\beta$) plane  
with seven different values of the $H^{\pm}$ mass, varying
from $300 \, \GeV$ to $1\, \TeV$ and nine values of $\tan\beta$ 
between $0.5$ and $50$. The mass of the pseudoscalar mediator $m(a)$ is set to $150 \,\GeV$ for this grid.
An additional scan of the pseudoscalar mediator $m(a)$ between $50\,\GeV$
and $375\,\GeV$ is performed, taking $m(H^{\pm}) = 500\,\GeV$ and
$\tan\beta = 1$, in order to assess the dependence of the results on
the $m(a)$ assumption. 
In both grid scans, the heavy scalar and pseudoscalar masses are always set to the same value
$m(H) = m(A) = m(H^{\pm})$.

\subsection{Background generation}
\label{sec:backgroundgeneration}

In order to describe the $t + \etmiss$ backgrounds accurately,  SM
processes involving at least one lepton coming from the decay of  vector
bosons are generated. 
Backgrounds either with fake electrons from jet misidentification
or with real non-isolated leptons  from the decay of heavy flavours are
not considered in our analysis, as a reliable estimate of these backgrounds
would require a simulation of detector effects beyond the scope of this
work. Based on ATLAS experimental results \cite{Aaboud:2017rzf}, we estimate
these backgrounds not to exceed around $15\%$ for the selections considered in
this paper. The backgrounds from $\ttbar$~\cite{Campbell:2014kua},
$tW$~\cite{Re:2010bp}, $WW$, $WZ$ and $ZZ$ production~\cite{Melia:2011tj,Nason:2013ydw}
were all generated at next-to-leading order (NLO) with {\tt POWHEG~BOX} \cite{Alioli:2010xd}. The
${\rm jets}+Z$ and ${\rm jets}+W$ samples are generated at~LO with 
{\tt  MadGraph5\_aMC@NLO}
and considering up to four jets for the matrix element calculation. 
{\tt  MadGraph5\_aMC@NLO} is also used to
simulate the $\ttbar V$ backgrounds with $V = W,Z$ at LO with a multiplicity
of up to two jets, and the $tZ$ and $tWZ$ backgrounds at LO.
The samples produced with {\tt POWHEG~BOX} are normalised to the NLO cross
section given by the generator, except $t\bar{t}$ which is normalised to
the  cross section obtained at next-to-next-to-leading order (NNLO)
plus next-to-next-to-leading logarithmic
accuracy~\cite{Czakon:2011xx,Czakon:2013goa}. The ${\rm jets} + W/Z$ samples
are normalised to the known NNLO cross sections~\cite{Anastasiou:2003ds,Gavin:2012sy},
and finally  the NLO cross sections calculated
with {\tt  MadGraph5\_aMC@NLO} are used as normalisations for the~$\ttbar V$~samples .  

\subsection{Detector smearing}
\label{sec:detectorsmearing}

Muons, electrons, photons, jets and \etmiss\ are constructed from the 
the stable particles in the generator output.
Jets are constucted by clustering the true momenta of all the
particles interacting in the calorimeters, with the exception of
muons. An 
anti-$k_t$ algorithm~\cite{Cacciari:2008gp}
with a parameter $R=0.4$ is used, as implemented in  {\tt FastJet}~\cite{Cacciari:2011ma}.
Jets originating from the hadronisation of bottom-quarks ($b$-jets) are experimentally
tagged with high efficiency (\btagged\ jets).
The variable \ptmiss \ with magnitude~\etmiss is defined at truth level,~i.e.~before
applying detector effects, as the negative of the vector sum of
the \pt s of all the invisible particles (neutrinos and DM
particles in our case). The effect of the detector on the kinematic quantities
utilised in the analysis is simulated by applying a Gaussian smearing to the momenta
of  the different reconstructed objects and reconstruction and tagging efficiency factors.
The parametrisation of the smearing and the reconstruction and tagging efficiencies is tuned to mimic
the performance of the  ATLAS detector~\cite{Aad:2008zzm,Aad:2009wy}
and is defined as a function of momentum and pseudorapidity  of the objects.
The discrimination of the signal from the background is greatly affected by the
experimental smearing assumed for the \etmiss, which is the main handle to tame
 the large~$\ttbar$~background. To this aim, the transverse momenta
 of unsmeared electrons, muons and jets are subtracted from the truth \etmiss
and replaced by the corresponding smeared quantities.  The residual truth
imbalance  is then smeared as a function of the scalar sum of the transverse
momenta of the particles not assigned to jets or electrons. 
The final selections and results are 
derived by analysing the simulated sample using the TDataFrame tool \cite{enrico_guiraud_2017_260230}.

\section{Kinematic properties of DM$t$ and analysis strategy}
\label{sec:analysis}

\begin{figure}[t]
\begin{center}
\includegraphics[width=0.48\textwidth]{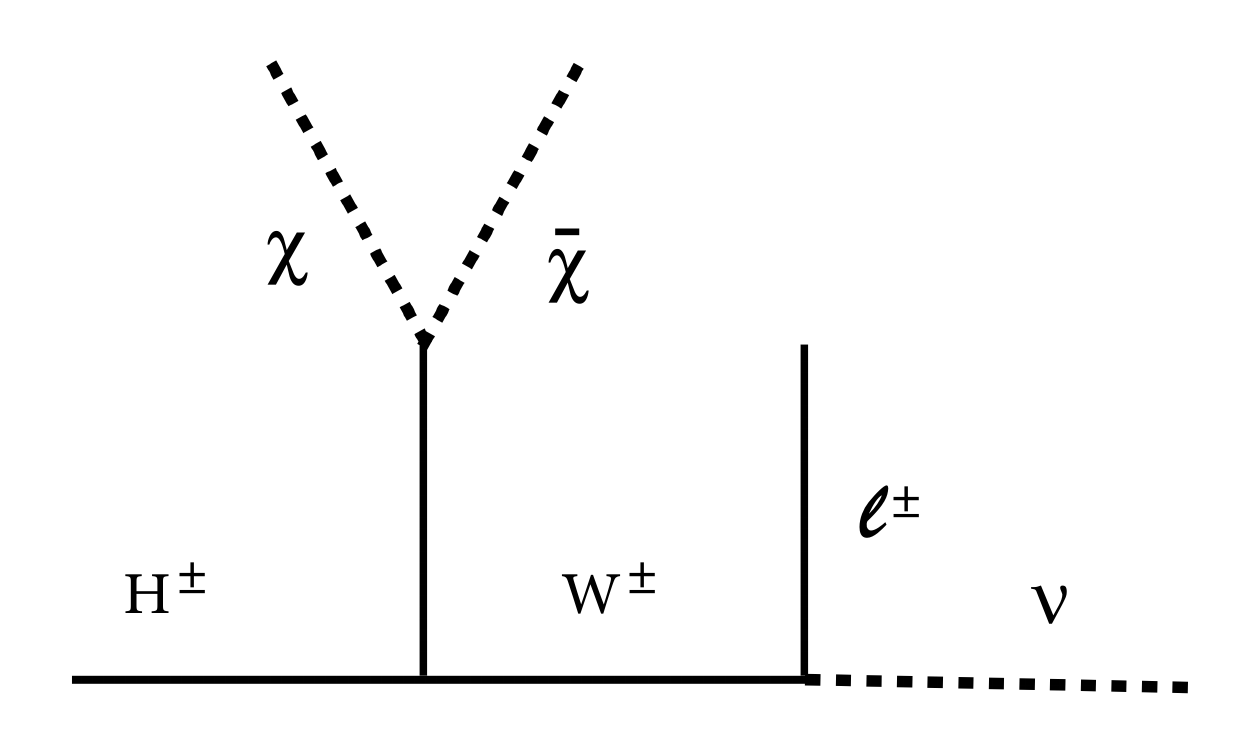}
\caption{Representation of the $H^{\pm}$ decay chain.}
\label{fig:kine}
\end{center}
\end{figure}

\begin{figure*}
\centering
\begin{subfigure}{.49\textwidth}\centering
\includegraphics[width=.9\textwidth]{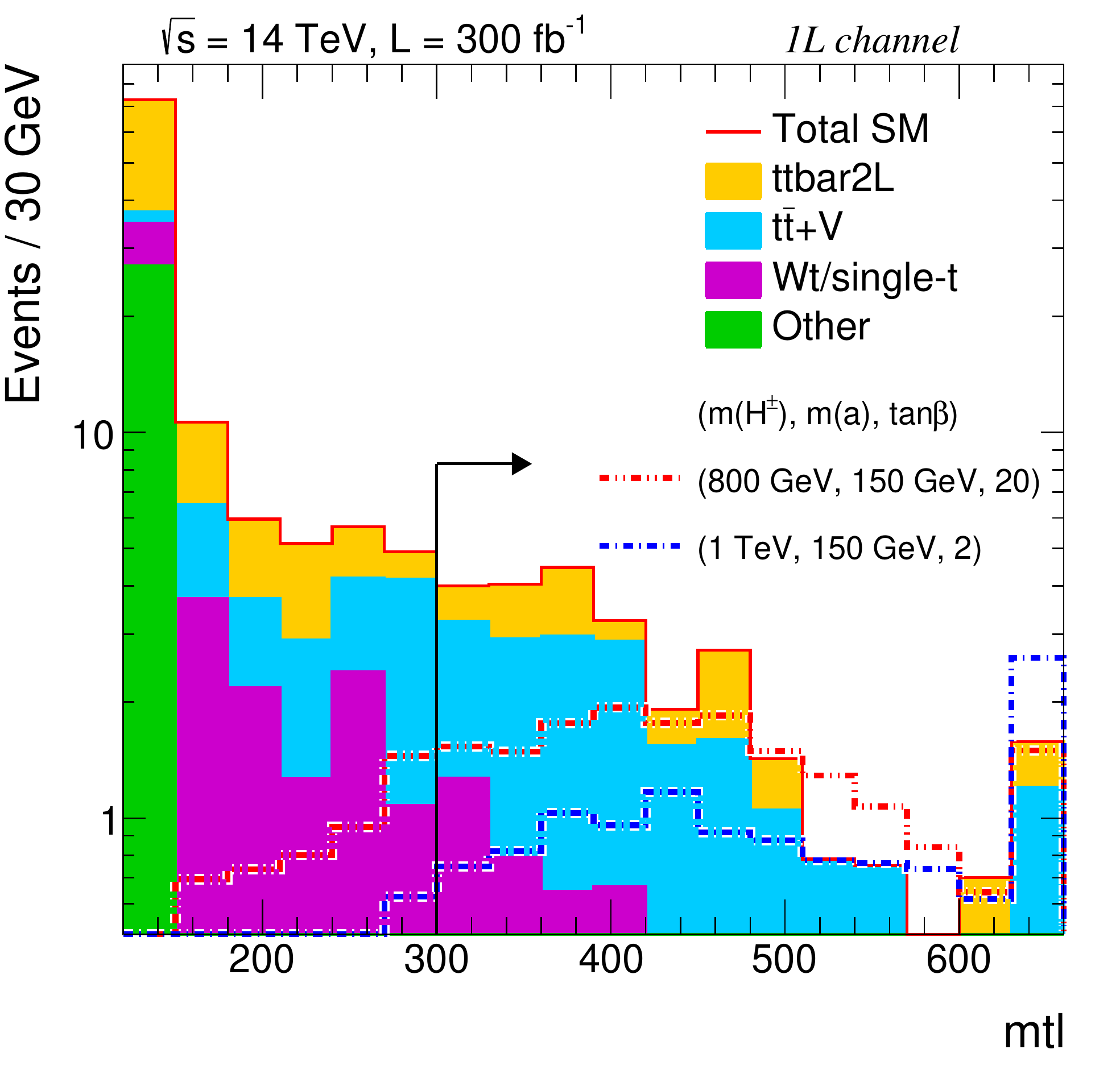}
\caption{}
\label{fig:kinematicsa}
\end{subfigure}
\begin{subfigure}{.49\textwidth}\centering
\includegraphics[width=.9\textwidth]{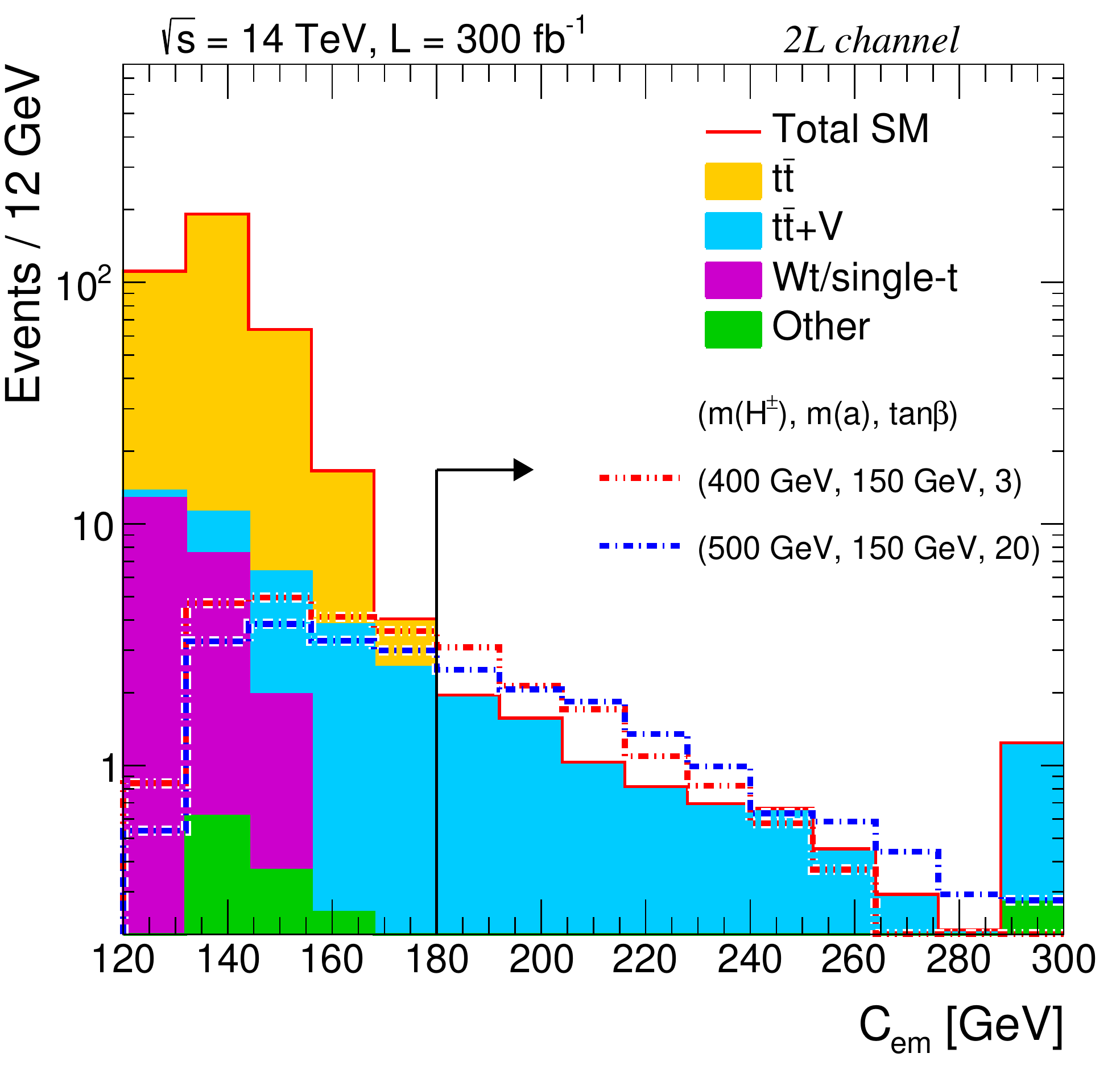}
\caption{}
\label{fig:kinematicsb}
\end{subfigure}
\caption{Distribution of the transverse mass variables used in the (a) one-lepton and (b) two-lepton selections after all requirements described in Sec.~\ref{sec:analysis}, except for the one on the plotted variable which is indicated with an arrow instead. The expected SM backgrounds and two signal benchmarks are compared in the figure for an integrated luminosity of 300~fb$^{-1}$ at the $14 \, \TeV$ LHC.}
\label{fig:kinematics}
\end{figure*}

The discussion of the  DM$t$ signal in Section \ref{sec:2HDM} should have made clear 
that the $tW$ channel
is the dominant production mechanisms for all parameter choices in which the $H^{\pm}$ can
decay on-shell into the pseudoscalar mediator and  a $W$ boson.
In order to search for this signal, we consider  two different final states in our analysis,
containing either one or two leptons.
In both cases the leptons are produced in the decay of a $W$ boson,
either prompt or in a cascade from the top-quark decay.
Furthermore, the signal events contain one $b$-jet,
which again stems from the top-quark decay.
In the one-lepton final state, two additional jets are produced from the hadronic
decay of one of the $W$ bosons.  A significant
amount of  \etmiss  associated to both the DM particles and the neutrinos
from  $W \to \ell \nu_\ell$ decays is also present in the events.

If the $W$ boson from $H^{\pm}\rightarrow aW^{\pm}$ decays 
leptonically into an electron or a muon, the resulting 
final state includes one lepton and three invisible particles,
with two invisible particles upstream of the lepton, and one downstream.
See Figure~\ref{fig:kine} for the corresponding decay chain.
The kinematics of this decay topology is analysed in the appendix of
Ref.~\cite{Polesello:2009rn}. The transverse mass 
$\mtlep$\footnote{%
We define:
$\mtlep = M_{\mathrm{T}}(\ptl,\ptmiss)^2\equiv2|\ptl||\ptmiss|(1-\cos\Delta\phi_{\ptl\ptmiss})$}
built with the components transverse to the beam of the lepton 
momentum ($\ptl$) and of the vector sum of the momenta of the 
invisible particles ($\ptmiss$) has  a distribution with an end-point
which is a function  of $m(H^{\pm})$, $m(W)$ and $m(a)$.
This variable can be directly measured for the one-lepton 
final state when the lepton is produced in the $H^{\pm}$ decay.
In the case of the  two-lepton final state, the distribution of $\mtlep$ for the $H^{\pm}$ 
decay enters the construction of the $\mttwo$ variable \cite{Lester:1999tx,Barr:2003rg} built 
out of the two leptons and $\ptmiss$, which has the same end-point. 

\begin{figure}[b]
\begin{center}
\includegraphics[width=0.49\textwidth]{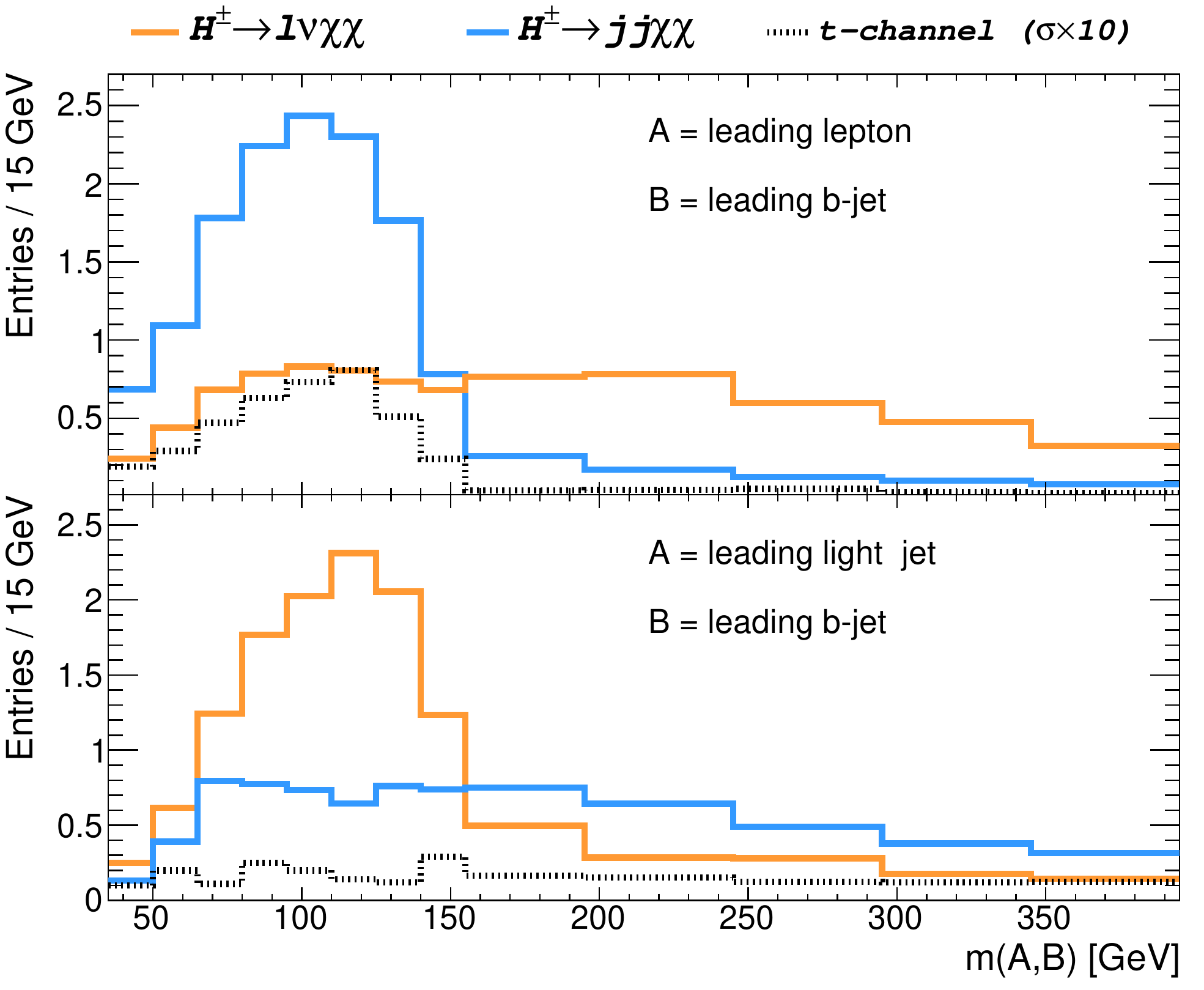}
\caption{The invariant mass of the lepton and the
leading \btagged\ jet ($m(b1,\ell)$) and of
the leading light jet and the leading \btagged\ jet ($m(b1,j1)$) are displayed for the lepton and hadronic decays of the $H^\pm$ in the $tW$ channel. 
For comparison also the distributions for $t$-channel production are shown. All results correspond to  $m(H^{\pm})=800 \, \GeV$ and $\tan\beta = 20$.}
\label{fig:Hleptruth}
\end{center}
\end{figure}

The  two variables $\mtlep$ and $\mttwo$ can therefore be exploited to strongly reduce the 
dominant SM backgrounds from single or double production of
top quarks, in which case $\etmiss$ is generated only from the neutrinos
from $W$ decay and the distributions display an end-point at the $W$-boson 
masses. 
The discriminating power of the \mtlep\ (left panel) and $\mttwo$ (right panel) observables
is illustrated in Figure~\ref{fig:kinematics}, which shows the relevant distributions 
after all the one-lepton and two-lepton signal selection
requirements, as described in the next section, have been applied.

The events surviving the \mtlep\ requirement for the one-lepton selection are dominated
by $t\bar{t}$ events that decay into two leptons, but one 
of the two leptons is not identified in the detector.
The variable $\amttwo$ \cite{Konar:2009qr,Lester:2014yga} was developed to 
tame this background and is therefore employed in the one-lepton analysis.

Secondary backgrounds like the production of single or double
vector bosons are efficiently suppressed by requiring QCD jet production
in the event. The angular 
correlation between the jets and $\etmiss$ has a good discrimination 
power, both for the 1-lepton and the 2-lepton case, as for 
leptonic decays of the $H^{\pm}$ all jets in the event are 
produced in the decay of the accompanying top quark, while 
$\etmiss$ is dominantly aligned with the $H^{\pm}$, implying
that  $\etmiss$ is mostly isolated from jets.

\subsection{One-lepton analysis}

Events with exactly one isolated lepton ($e$ or $\mu$) with 
$\pt > 25\,\GeV$, at least three jets
($\pt > \{50,50,20\}\,\GeV$, $|\eta|<2.5$) and $\etmiss > 250\,\GeV$ are
selected for this topology.  
All reconstructed jets with $p_\mathrm{T}^j >25 \, \GeV$ 
within $|\eta_{\ell}|<2.5$ have to satisfy $|\Delta\phi_{\rm min}|>1.1$, 
where~$\Delta\phi_{\rm min}$ is defined to be the angle between 
$\vec{p}_{\rm T}^{\ j}$ and $\ptmiss$ for the jet closest to $\etmiss$ 
in the azimuthal plane. At least one jet is required to be \btagged.  
All dominant backgrounds except for single top production are
characterised by two hard \bjet s produced in the decay of a top
quark.  
In order to suppress these backgrounds, event with a second
\btagged\ jet with $\pt > 50\,\GeV$ are rejected. The semi-leptonic
and dileptonic \ttbar\ backgrounds are strongly suppressed by
requiring $\mtlep > 300\,\GeV$ and $\amttwo > 200 \,\GeV$,
respectively.

Further requirements on the invariant mass of the lepton and the
leading \btagged\ jet ($m(b1,\ell) > 160\,\GeV$) and on the invariant mass of
the leading light jet and the leading \btagged\ jet ($m(b1,j1) < 150\,\GeV$)
are placed to further suppress the residual background compatible with
the presence of a semileptonic top decay in the event. As it can be
seen in Figure~\ref{fig:Hleptruth}, these requirements select the signal
topology where the $H^{\pm}$ decays leptonically, which was found to
have kinematic features that made it easier to discriminate it
from the backgrounds.  
The signal events where the $H^{\pm}$  decays hadronically are
kinematically more similar to the SM backgrounds, due to the smearing
of \etmiss\ associated to neutrinos in top decays.
In this case  a dedicated strategy would be needed to successfully
distinguish between signal and background.  The same applies to the
production via $t$-channel diagrams, which is also rejected by the requirements
of the analysis. The definition of a dedicated signal region
targeting the hadronic $H^{\pm}$ decay is expected to increase
significantly the sensitivity of the analysis. We leave the definition
of such region to the experimental collaborations.


\subsection{Two-lepton analysis}

As a first step, events with two leptons and at least one \btagged\ jet
are selected. The events are required to contain
exactly two isolated oppositely charged  leptons (electrons,
muons or one of each flavour) with $p_\mathrm{T}^{\ell_1} >25 \, \GeV$,
$p_\mathrm{T}^{\ell_2} >20 \, \GeV$, $|\eta_{\ell}|<2.5$ and an  invariant
mass that satisfies  $m_{\ell \ell} > 20 \, \GeV$.
If the charged signal leptons are of the same flavour
the additional requirement $m_{\ell \ell} \in [71, 111] \, \GeV$ is
imposed to veto events where the charged lepton pair  arises from a~$Z
\to \ell^+ \ell^-$ decay. Furthermore, each event is required to 
contain at least one \btagged\ jet with $\pt >40\,\GeV$.
All reconstructed jets with $\pt^j >25 \, \GeV$ 
within $|\eta_{\ell}|<2.5$ have to satisfy $|\Delta\phi_{\rm
  min}|>1.5$. The variable $\Delta\phi_{\rm boost}$, the azimuthal angular distance between $\ptmiss$ and the vector sum of $\ptmiss$ and the transverse 
momenta of the leptons must satisfy the requirement $|\Delta\phi_{\rm boost}|<1$.
The reducible backgrounds are suppressed by requiring that the
invariant mass of at least one lepton with the leading $b$-jet is
smaller than $150\,\GeV$, and thence compatible with the decay of a top quark.
The dominant $t\bar{t}$ backgrounds have a second \btagged\ jet, 
with $p_T$ typically in excess of $50\,\GeV$, whereas the signal 
has only one top decay. The requirement that the scalar sum 
of the transverse momenta of all the jets observed in the 
event be lower than $150\,\GeV$ suppresses events with two 
real top quarks.
The final cut, following \cite{Haisch:2016gry} is based on the following  
linear combination of~$\etmiss$ and~$\mttwo$: 
\beq \label{eq:Cem}
C_{\rm em} \equiv \mttwo + 0.2 \cdot \etmiss \,.
\eeq
The requirement that this variable be larger than $180\,\GeV$, together
with the cut $\mttwo>100\,\GeV$ reduces the background from
$t\bar{t}$ production well below the irreducible $t\bar{t}+Z$ background.
This is shown in the right panel of Figure~\ref{fig:kinematics}.



\section{Results}
\label{sec:results}

On the basis of the selection criteria defined in the previous section,
we study the LHC sensitivity to the DM$t$ signature  for
an integrated  luminosity of $300 \, {\rm fb}^{-1}$ at $\sqrt{s} = 14 \, \TeV$. 

The total background in the one-lepton selection is approximately 25 events.
More than half of the background contribution is coming from $tt+V$ and $tZ$ processes
and the rest is due to the contribution of top pairs (dileptonic decays) and single top $tW$ channel in an approximate ratio of 2 to 1.
In the charged Higgs mass range from 
$500 \, \GeV $ to $1\, \TeV$ the acceptance for signal events containing at 
least one lepton
amounts to  $[0.5, 1]\%$ ($[0.2, 0.8]\%$) for $m(a) = 150\, \GeV$ and 
$\tan\beta = 1 \, (20)$.
The total background in the two-lepton selection is approximately 10 events, dominantly composed of
the $\ttbar+V$ and $tWZ$ background processes.
For $m(H^\pm)$ between $300 \, \GeV$ to $700\, \GeV$ the acceptance 
for signal events containing at least two leptons
is in the range $[0.1, 0.7]\%$ ($[0.06, 0.5]\%$) for $m(a) = 150\, \GeV$ and 
$\tan\beta = 1 \, (20)$.

A profiled likelihood ratio test statistic is used to evaluate the
upper limit on the ratio of the signal yield to that predicted in the
2HDM+$a$ model. The CLs method \cite{Read:2002hq} is used to derive exclusion limits at 95\%
Confidence Level (CL). The statistical analysis has been performed
by employing the RooStat toolkit \cite{Moneta:2010pm}.
The results are interpreted in terms of relevant parameters defining the model, namely
$m({H^\pm})$,  $m(a)$ and $\tan \beta$.
The masses of the other Higgs bosons, except for the SM one, are set to the mass of the charged Higgs. 

\begin{figure}[h]
\begin{center}
\begin{subfigure}{.49\textwidth}\centering
\includegraphics[width=.98\textwidth]{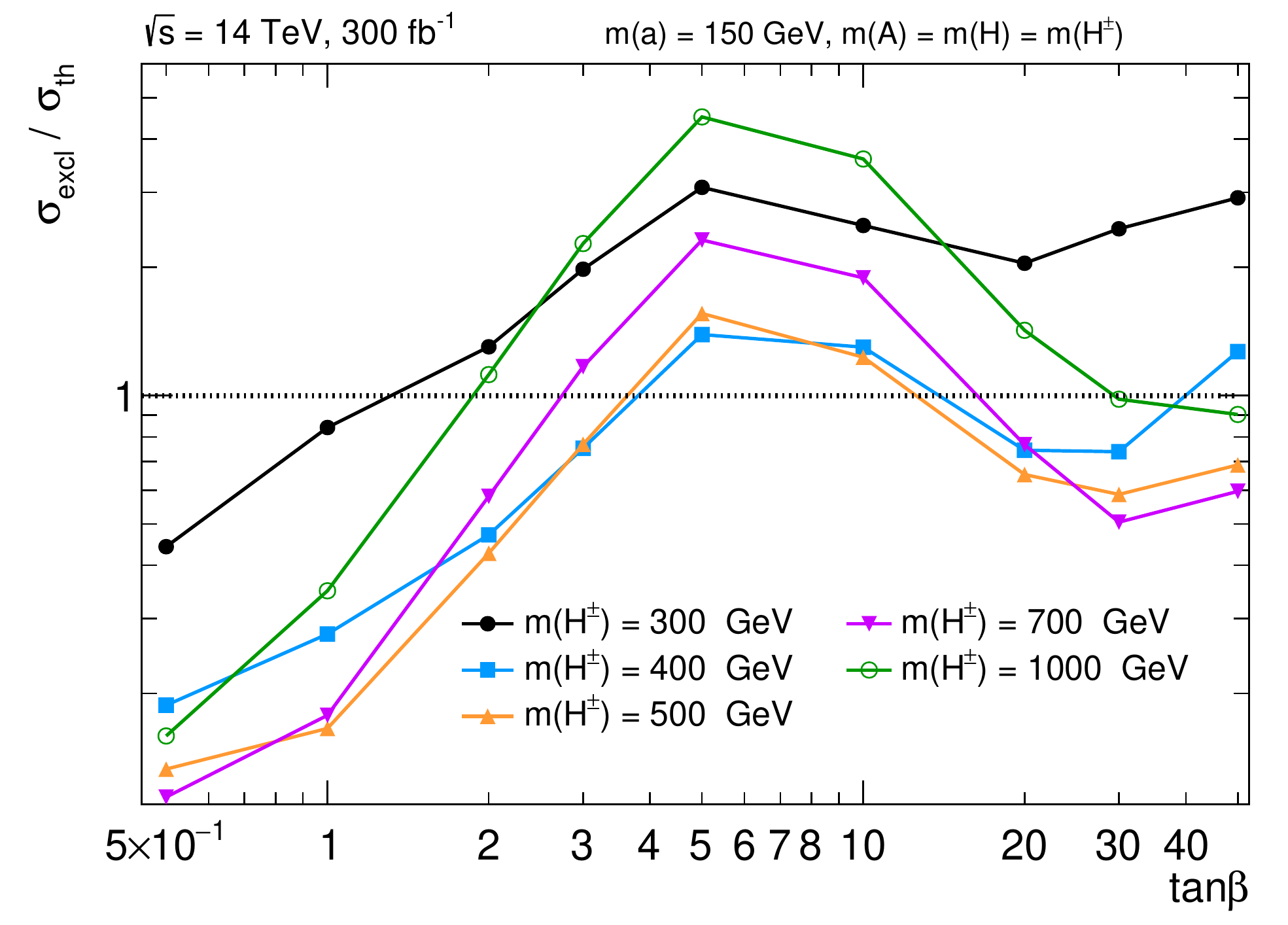}
\caption{}
\label{fig:95clSpina}
\end{subfigure}
\begin{subfigure}{.49\textwidth}\centering
\includegraphics[width=.98\textwidth]{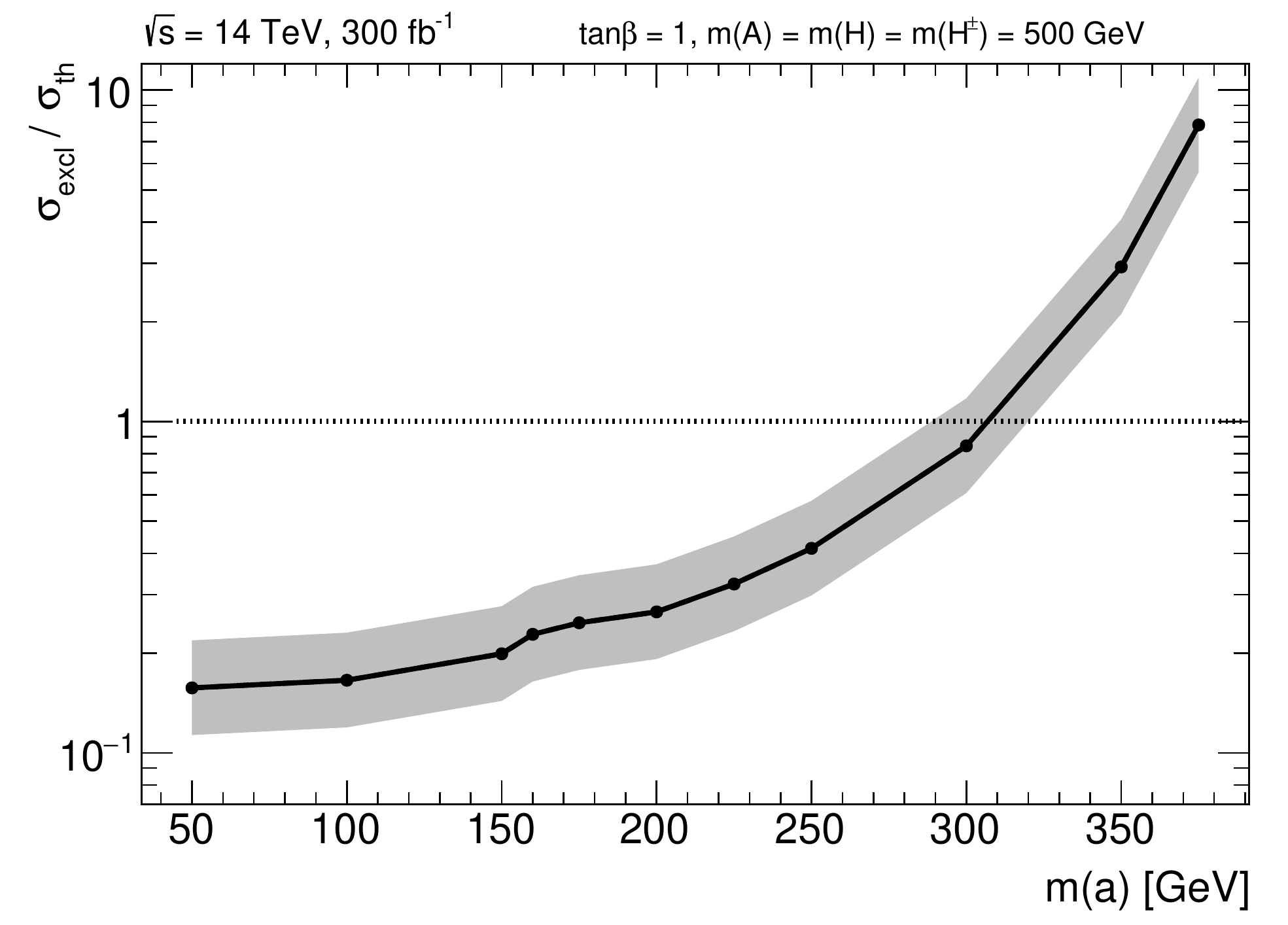}
\caption{}
\label{fig:95clSpinb}
\end{subfigure}
\caption{Upper limits at 95\% CL  on the ratio of the signal yield to that predicted in the
2HDM+$a$ model using the combination of the one-lepton and two-lepton selections described in the text.
The limits are presented in (a) as a function of $\tan\beta$ for different $m(H^\pm)$ masses and $m(a) = 150\,\GeV$, and in (b) as a function of  $m(a)$
for  $m(H^\pm) = 500\,\GeV$ and $\tan\beta = 1 $.
The reach assumes $300 \, {\rm fb}^{-1}$ of $14 \, \TeV$ LHC data and a systematic  uncertainty of 20\% (5\%) on the SM background (signal).}
\label{fig:95clSpin}
\end{center}
\end{figure}

Given the relatively large irreducible background surviving all the
selections, the experimental sensitivity will be dominantly  determined by
the systematic uncertainty on the estimate of the SM backgrounds. Such
uncertainty has  two main sources: the uncertainties affecting 
the detector performance such as the energy scale 
for hadronic jets and the identification efficiency for leptons, 
and, in addition, the uncertainties plaguing 
the evaluation procedure for the background which typically 
includes a mix of theoretical uncertainties on 
the MC modelling of SM processes and uncertainties on 
the data-driven estimates of the main backgrounds.  
Depending on the process and on the kinematic selection, 
the total uncertainty can vary between a few
percent and a few tens of percent. Since the present analysis does not select
an extreme kinematic phase space for the dominant $\ttbar Z$ background,
it should be possible to control the systematic uncertainties
at the 10\% to 30\% level.  In the following, we will assume a 20\%
uncertainty on the backgrounds and, furthermore,  a 5\% uncertainty on the
signal, which accounts for the impacto of scale and PDF variations on the signal modelling.

\begin{figure}
\begin{center}
\includegraphics[width=0.5\textwidth]{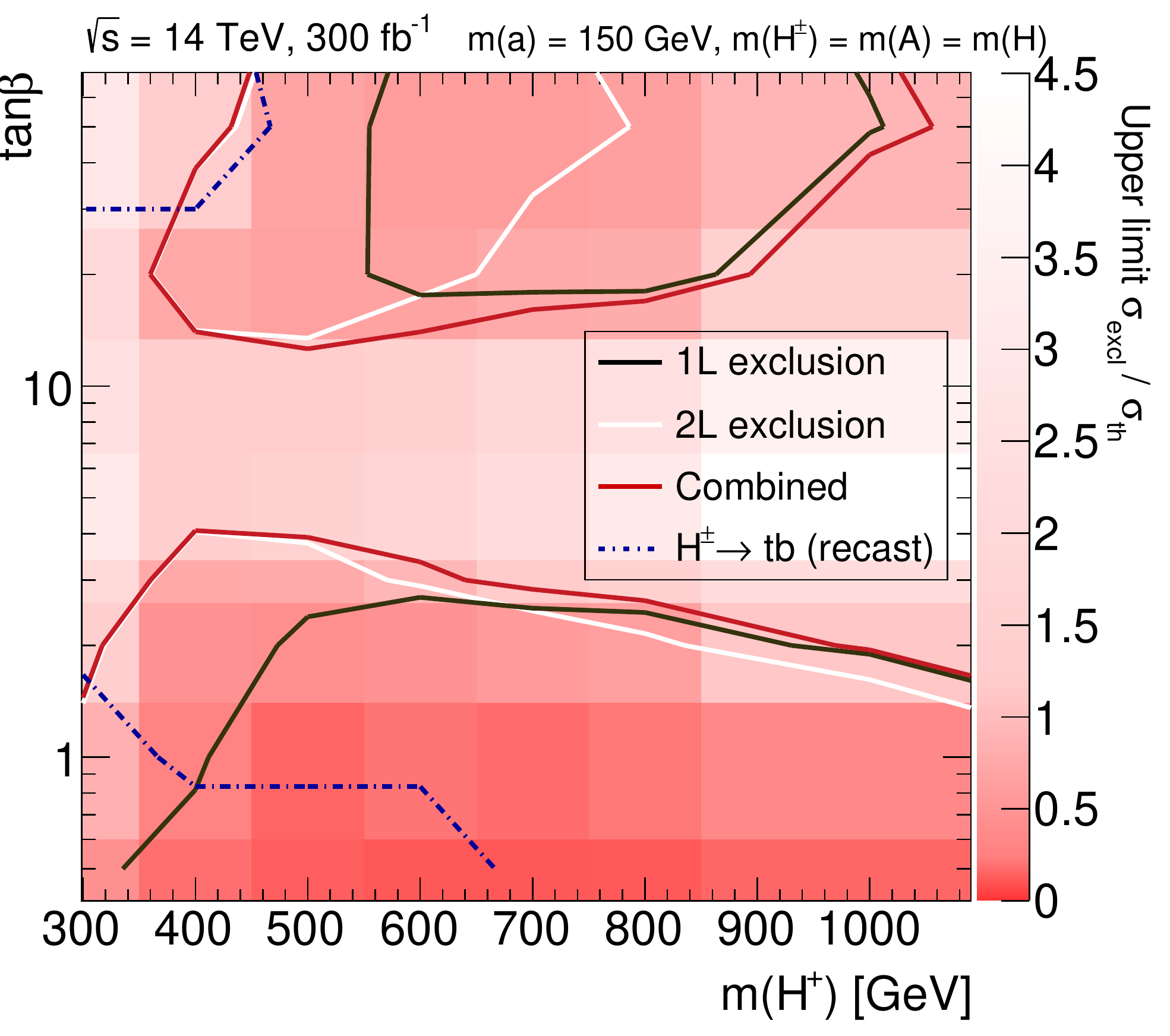}
\caption{Regions in the (m($H^{\pm}$), $\tan\beta$) plane which can be excluded at 95\%~CL through the one-lepton and two-lepton searches described in the 
text. The reach assumes $300 \, {\rm fb}^{-1}$ of $14 \, \TeV$ LHC data and a systematic  uncertainty of 20\% (5\%) on the SM background (signal).}
\label{fig:95clSpinc}
\end{center}
\end{figure}

Since the one-lepton and two-lepton analyses select two orthogonal event 
samples, they can be statistically combined, in order to assess the 
potential gain in sensitivity deriving from such treatment. In the combination, 
both signal and background uncertainties are treated as correlated. 

Figure~\ref{fig:95clSpina} shows the exclusion limits obtained by the
combination of the one-lepton and two-lepton selections for different charged
Higgs masses as a function of $\tan\beta$. The sensitivity trend
closely follows the cross-section distribution shown in Figure~\ref{fig:xstb}.
The maximum of the sensitivity is found for $m(H^{\pm}) = 500\,\GeV$, 
while $\sigma_{\rm excl}/\sigma_{\rm th}$ is relatively
flat for masses between $400 \, \GeV$ and $700\, \GeV$.
In Figure~\ref{fig:95clSpinb} we instead show the exclusion limits as a
function of the light pseudoscalar mass for $m(H^{\pm})$ and $\tan\beta$ 
set to $500\,\GeV$ and $1$ respectively. One observes that the sensitivity 
is relatively flat for
$m(a)$ values between $50 \, \GeV$ and $200\, \GeV$, and 
that for  the chosen parameters $\sigma_{\rm excl}/\sigma_{\rm th} < 1$ for 
$m(a) \lesssim 300 \, \GeV$.

Finally, Figure~\ref{fig:95clSpinc} shows the exclusion contour in the
$(m(H^{\pm}), \tan\beta)$ plane for the separate one-lepton and two-lepton
selections and their combination. The $z$-axis shows the ratio of the
excluded and the theoretical cross-sections for the combined fit. 
Comparing the contours,  the complementarity
in reach between the one-lepton and the two-lepton selections is evident, 
resulting only in a small improvement when  the two channels are combined.

Limits on  the production of $H^{\pm}$ followed by the decay into either 
$\tau\nu_{\tau}$ \cite{Aaboud:2016dig,CMS-PAS-HIG-16-031} or $tb$ 
\cite{Aad:2015typ,Khachatryan:2015qxa,ATLAS-CONF-2016-089} are available  from the ATLAS and CMS collaborations. 
For the decay into $\tau\nu$ the limits are outside 
the range of parameters considered in this analysis.
For the $tb$ decay we recast the limits given in \cite{ATLAS-CONF-2016-089} 
taking into account  
that the $H^{\pm} \to tb$ BR is reduced because the partial decay width
$H^{\pm}\rightarrow a W^{\pm}$ is in general non-vanishing in our model~(\ref{eq:VP}).
The results are shown as a blue dashed line in Figure~\ref{fig:95clSpinc}, and 
they cover an area largely complementary to the results  of the DM$t$ analysis.

\section{Conclusions}
\label{sec:conclusions}

In this article we have assessed the prospects of future LHC runs to probe
spin-0 interactions between DM and top quarks via the $t + \etmiss$
signature. We have focused on a model 
with two Higgs doublets and a pseudoscalar mediator.
The rich structure of the Higgs sector in the the 2HDM+$a$ model provides 
interesting final-state signatures dependent on the mass hierarchy
of the different bosons. In particular, the $t + \etmiss$ signature is 
dominated by on-shell production of the charged Higgs
associated with a  top-quark, if the decay channel  
$H^{\pm} \rightarrow W^{\pm} a$ is kinematically accessible.

Two final states were considered, involving one and two 
leptons from the decay of the two $W$ bosons in the event.
Analysis strategies were developed which take advantage of the 
topology of the leptonic $H^{\pm}$ decay to enhance the signal
with respect to the SM backgrounds. It was shown that the one-lepton 
and two-lepton analyses have complementary sensitivity as 
a function of $m(H^{\pm})$, with the former (latter) being more sensitive 
at higher (lower)  masses.

For a mediator with mass $150 \, \GeV$, and maximally mixed with the 
pseudoscalar $A$ of the two Higgs doublet model, 
values of  $\tan\beta$ up to 3 and down 
to 15 can be excluded at 95\%~CL by the LHC
with an integrated luminosity
of 300~fb$^{-1}$  at $\sqrt{s} = 14 \, \TeV$, 
if the $H^{\pm}$ mass is in the range of $300 \, \GeV$ and $1 \, \TeV$.
The  $t + \etmiss$ signature considered here for the first time  
therefore complements the parameter coverage of the
mono-Higgs, mono-$Z$ and $t\bar t +\etmiss$ searches that have been discussed 
in the context of the  2HDM+$a$ model in~\cite{Bauer:2017ota}.

\section{Acknowledgments}
We thank Uli Haisch and Fabio Maltoni for several enlightening 
conversations  during the development of the analysis, and for useful 
comments  on the manuscript. This work was started during the 
BSM session of the workshop "Physics at TeV Colliders"  
held in Les Houches in June
2017. We thank the organisers for the stimulating
environment of the workshop, and the colleagues taking part 
in the meetings for useful discussions.



\bibliographystyle{elsarticle-num} 
\bibliography{hplus}
\end{document}